\documentclass[aps,prd,preprint,groupedaddress,amssymb]{revtex4}
\usepackage{graphics}

\def\plb#1{Phys.~Lett.~{\bf B#1}}

\def\prl#1{Phys.~Rev.~Lett.~{\bf #1}}
\def\prd#1{Phys.~Rev.~{\bf D#1}}

\def\ptp#1{Prog.~Theor.~Phys.~{\bf #1}}

\begin{document}

\preprint{UCD-03-13}

\title{Bi-large Neutrino Mixing and Mass of the Lightest Neutrino from Third Generation Dominance in a Democratic 
Approach}

\author{Radovan Derm\' \i\v sek}
\email[]{dermisek@physics.ucdavis.edu}

\affiliation{Davis Institute for High Energy Physics,
University of California, Davis, CA 95616, U.S.A.}

\date{October 1, 2004}

\begin{abstract}

We show that both small mixing in the quark sector and large mixing in the
lepton sector can be obtained from a simple assumption of universality of Yukawa couplings and the right-handed 
neutrino Majorana mass matrix in leading order.
We discuss conditions under which bi-large mixing
in the lepton sector is achieved with a minimal amount of fine-tuning requirements for possible models.
From knowledge of the solar and atmospheric mixing angles we determine the allowed values of $\sin \theta_{13}$.
If embedded into grand unified theories, the third generation Yukawa 
coupling unification is a generic feature while masses of the first two generations of charged fermions depend on 
small perturbations. In the neutrino sector, the heavier two neutrinos are model dependent, while the mass of the 
lightest neutrino in this approach does not depend on perturbations in the leading order.
The right-handed neutrino mass scale can 
be identified with the GUT scale in which case
the mass of the lightest neutrino 
is given as $(m_{top}^2/M_{GUT}) \sin^2 \theta_{23} 
\sin^2 \theta_{12}$ in the limit $\sin \theta_{13} \simeq 0$.
Discussing symmetries 
we make a connection with hierarchical models and show that the basis independent characteristic of this scenario is 
a strong dominance of the third generation right-handed neutrino, $M_1, M_2 < 10^{-4} M_3$, $M_3 = M_{GUT}$.

\end{abstract}

\pacs{}

\maketitle




\section{Introduction}

The masses of three generations of fermions in the standard model are 
scattered in six orders of magnitude between the mass of the electron (0.5 
MeV) and the mass of the top quark (175 GeV). Neutrino experiments suggest 
that  masses of neutrinos ($< 0.05$ eV) are another seven orders of 
magnitude 
lighter than the electron. Furthermore, the mixing angles in the 
quark sector given by the Cabibbo-Kobayashi-Maskawa matrix $V_{CKM}$ are 
small, while the mixing in the lepton sector is large.   
The solar neutrino data point to large mixing and the atmospheric neutrino 
data require close to maximal mixing.
This is one of the most challenging puzzles in elementary particle physics.

A lot of effort was made in order 
to understand the origin of hierarchy and mixing in fermion 
masses~\cite{Fritzsch_Xing_review}.
Some of the most promising models 
are based on grand unified theories [GUTs] in which 
all fermions originate from just a few multiplets of GUT gauge 
symmetry group. 
Putting all particles of one generation into the same multiplet (as it is, for example, 
in SO(10)) offers a very attractive 
possibility that their Yukawa couplings are all equal at the GUT scale in a similar way 
to the well-established gauge coupling unification. This works quite well for the third 
generation of fermions but fails for the first two generations. 
Explaining the spectrum 
of the first two generations requires very specific assumptions about the underlying 
theory at the GUT scale. 

In the hierarchical approach, 
the unification of the third generation Yukawa couplings
is typically the starting point 
and in the leading order only this coupling is generated. This is usually achieved by imposing 
family symmetries. Yukawa couplings of the first two families are generated in the process of family symmetry 
breaking.
Although these models can reproduce the observed spectrum and mixing,
the level of model building involved
casts a shadow on third generation Yukawa 
unification itself. Why should the third generation be special when it is so easy to build 
models which do not unify the first two?
Small mixing angles in this approach can be understood as a  consequence of the large 
hierarchy in mass matrices. However, in order to obtain large mixing in the lepton sector it 
is often assumed 
that neutrinos are very different; the neutrino Yukawa matrix {\it is not} hierarchical or the right-handed 
Majorana mass matrix has a special form~\cite{Fritzsch_Xing_review,neutrino_reviews}, or 
the neutrino sector is completely random~\cite{anarchy}.

Hierarchy in quark masses can also be  understood within a democratic 
approach~\cite{democracy}, in which all Yukawa couplings are identical 
in the leading order 
and all differences result from small departures from universality.
However, in order to generate large mixing in the 
lepton sector, neutrinos are again assumed to be special. This time it is required that the 
neutrino mass matrix {\it is not} democratic (it is diagonal or it has another special form). 
This 
is either a by hand assumption~\cite{dem_neutrino} or it can be achieved in specific 
models~\cite{Mohapatra_Nussinov}.

In what follows we show that both small mixing in the quark sector and large mixing in the 
lepton sector can be obtained from a simple assumption of universality of Yukawa couplings. 
After introducing notation in Sec.~\ref{sec:notation}, we start with a discussion of two families only, 
Sec.~\ref{sec:2g}, since the mechanism which generates large mixing in the lepton sector is easier to follow
in this case.
Within a democratic approach,
assuming the same universal form of all Yukawa matrices {\it and} the right-handed neutrino 
Majorana mass matrix (in the leading order), we identify a condition under which large mixing 
in the lepton sector is achieved. 
We show that the universal part of the resulting left-handed neutrino mass matrix is washed 
out due to the seesaw mechanism and the dominant contribution to neutrino masses comes from small departures from 
universality. This is what distinguishes quarks from leptons.
The case of three families is discussed in detail in Sec.~\ref{sec:3g}.

The virtue of this approach is that all mass matrices are treated in the same way, providing a simple 
framework which can be easily embedded into more fundamental theories.
If embedded into GUTs, the third generation Yukawa
coupling unification is a generic feature in this approach, while the spectrum of the 
first two generations of quarks and charged leptons crucially depends on small perturbations. 
This becomes even more obvious in Sec.~\ref{sec:symmetries}, where symmetries of this framework are discussed and 
the connection with hierarchical models is made. It is shown that the distinguishing (basis independent) 
feature of this approach is the dominance of the third generation right-handed neutrino mass. 

Masses of light neutrinos do not follow the same pattern as masses of quarks and charged leptons.
The two heavier neutrinos are given in terms of perturbations and so are highly model dependent, while
the mass of the lightest neutrino does not depend on details of a model in the leading order.
If the right-handed neutrino scale is identified with the GUT scale, the mass of the lightest neutrino
is predicted and it is related to the elements of the lepton mixing matrix. Avoiding the necessity of introducing an 
intermediate scale for right-handed neutrinos makes this framework very 
predictive.
Finally, we conclude in Sec.~\ref{sec:conclusions}.

After the first version of this paper was finished~\cite{Dermisek:2003rw} it was brought to our attention that 
a democratic form of both 
Yukawa matrices and the right-handed neutrino Majorana mass 
matrix was previously suggested in~\cite{branco_efd}. Some of our results, 
namely the conditions under which large mixing in the lepton sector is 
achieved, coincide with the findings in Ref.~\cite{branco_efd}.
For related studies, see also ~\cite{branco_Z3, teshima}. 

\section{\label{sec:notation} Notation}

The masses of quarks and leptons originate from Yukawa couplings of matter fields to
one or more Higgs bosons. When Higgs fields acquire  vacuum expectation values [vevs], 
the Lagrangian containing mass terms of quarks and leptons can be written as:
\begin{equation}
{\cal L}_m = 
- v_f \bar{f}_{Li} (Y_f)_{ij}  f_{Rj}  + h.c. , \quad f = u, d, e, \nu ,
\end{equation}
where $v_f$ result from vevs of Higgs fields which couple to the corresponding quark or 
lepton,
$f_L$ ($f_R$) represent left-handed (right-handed) fields, 
 and
$Y_f$  are Yukawa couplings, in general arbitrary $3 \times 3$ complex 
matrices in generation space represented by subscripts $i,j$. 
The Yukawa matrices can be diagonalized by bi-unitary 
transformations:
\begin{equation}
\hat{Y}_f = U_f Y_f V_f^\dag , 
\end{equation}
where $\hat{Y}_f$ are diagonal matrices containing mass eigenvalues and $U_f$ and $V_f$ are 
unitary 
matrices. The mismatch between diagonalization of up-quark and down-quark Yukawa matrices 
appears in the charged current Lagrangian in the form of the CKM matrix:
\begin{equation}
V_{CKM} = U_u U_d^\dag .
\end{equation}

The smallness of neutrino masses can be  naturally explained by
the seesaw
mechanism~\cite{see-saw} which assumes the existence of  Majorana
masses for right-handed neutrinos:
\begin{equation}
{\cal L}_{\nu_R} = - \frac{1}{2} \nu_R^T M_{\nu_R} \nu_R + h.c.
\end{equation}
where $M_R$ is a matrix in generation space.
When right-handed neutrinos are integrated out we obtain a Majorana mass matrix for 
left-handed neutrinos:
\begin{equation}
M_{\nu_L} = - v^2_\nu Y_\nu M_{\nu_R}^{-1} Y_\nu^T ,
\label{eq:see-saw}
\end{equation}
which can be diagonalized by a single unitary matrix: 
\begin{equation}
\hat{M}_{\nu_L} = U_{\nu_L} M_{\nu_L} U_{\nu_L}^T .
\end{equation} 
And finally, the lepton mixing matrix which appears in the charged current Lagrangian is given as: 
\begin{equation}
U = U_e U_{\nu_L}^\dag.
\label{eq:U}
\end{equation}

\section{\label{sec:2g} Democratic matrices: two families}

Let us start with only two generations and let us assume their Yukawa couplings are universal 
in the leading order:
\begin{equation}
Y_f \approx \frac{1}{2} \lambda_f \ {\cal I}  , \quad
{\cal I} = 
                 \left(\begin{array}{cc}  1 & 1  \\
                                          1 & 1  \\ 
                 \end{array} \right) .
\end{equation}
If the Yukawa matrices are exactly equal to ${\cal I} \lambda_f/2 $, mass eigenvalues 
are $\{ 0,\lambda_f \}$ and the diagonalization matrix is:
\begin{equation}
U_{\cal I} = \left(\begin{array}{rr}
                            -\frac{1}{\sqrt{2}} & \frac{1}{\sqrt{2}} \\
                             \frac{1}{\sqrt{2}} & \frac{1}{\sqrt{2}} 
                   \end{array} \right)  .
\label{eq:UI}
\end{equation}
Therefore one generation is massless in the leading order and the CKM matrix is
the identity matrix as a consequence of the unitarity constraint $V_{CKM} = U_u U_d^\dag =
U_{\cal I} U_{\cal I}^\dag = 1 $. This is quite a good approximation to reality, 
taking 
into
account that the only assumption so far is the universality of Yukawa couplings.

Now let us parametrize the departure from universality by matrices ${\cal E}_f$ so 
that~\footnote{For simplicity we assume real symmetric Yukawa matrices.}:
\begin{equation}
Y_f \equiv \frac{1}{2} \lambda_f \left( {\cal I}  - {\cal E}_f \right)  , \quad
{\cal E}_f = \left(\begin{array}{cc}  \epsilon_{f 11} & \epsilon_{f 12}  \\
                                      \epsilon_{f 12} & \epsilon_{f 22}  \\ 
                 \end{array} \right) .
\end{equation}
Taking, for example, $\epsilon_{f 11} = 0$ and $\epsilon_{f 12} = \epsilon_{f 22} 
\equiv \epsilon_f $ 
we find $m_s/m_b \simeq \epsilon_d /4$,  $m_c/m_t \simeq \epsilon_u /4$ and 
$V_{cb}  
\simeq (\epsilon_d-\epsilon_u )/4 \simeq m_s/m_b$. 
The predicted value 
of $|V_{cb}|$ in this case is 0.02 which is not so far from the desired value $|V_{cb}| = 
0.036$.
Clearly, relaxing the condition between elements of $\cal E$, there is enough 
freedom to fit precisely both quark masses and CKM elements.
For specific models see~Ref.~\cite{Fritzsch_Xing_review}.
The generic 
consequence in this approach is that the value of $|V_{cb}|$ is proportional to 
the generated hierarchy in quark masses.
The diagonalization
matrices differ only by $O (\epsilon)$ from $U_{\cal I}$. Therefore the resulting CKM
matrix differs from the identity matrix by  $O (\epsilon)$ although the mixing in both up and 
down sectors is close to maximal. 


Let us turn our attention to the lepton sector. If the only source of neutrino mass was 
the neutrino
Yukawa matrix, the lepton mixing matrix would naturally be close to the identity matrix (in the 
same 
way as the CKM matrix), although it is reasonable to expect off-diagonal
elements to be much larger than the corresponding CKM elements due to the fact that the
hierarchy in both charged lepton and neutrino sectors is much smaller than in the quark
sector ($m_{\nu_2}/m_{\nu_3} \gtrsim 0.16$). However, close to maximal mixing would require 
fine-tuning ($\epsilon_{\nu 12} \sim 1 $). 
Fortunately, we will see that the lepton mixing matrix can naturally be very different 
when considering right-handed neutrino Majorana mass and the  
seesaw mechanism.

Let us make a similar assumption about the form of $M_{\nu_R}$ as we made for Dirac Yukawa 
matrices:
\begin{equation}
M_{\nu_R} = \frac{1}{2} \left( {\cal I} - {\cal R} \right) M_0 , \quad
{\cal R} = \left(\begin{array}{cc}  r_{11} & r_{12}  \\
                                     r_{12} & r_{22}  \\
                 \end{array} \right) .
\end{equation}
It is useful to write the inverse of this matrix as:
\begin{equation}
M_{\nu_R}^{-1} = \frac{1}{M_{eff}} \left( \hat{\cal I} + \hat{\cal R} \right),
\label{eq:Mr-1}
\end{equation}
where
\begin{equation}
\hat{\cal I} \equiv 
                 \left(\begin{array}{rr}
                                      -1 & \ \ 1  \\
                                       1 &    -1   \\
                 \end{array} \right) , \quad
\hat{\cal R} \equiv 
                 \left(\begin{array}{rr}
                                        r_{22} & - r_{12}  \\
                                    - r_{12}   &  r_{11}  \\ 
                 \end{array} \right) ,
\label{eq:hatIR}
\end{equation}
and the effective right-handed neutrino mass scale is
\begin{equation}
M_{eff} = \frac{1}{2} \left( r - \det{\cal R}  \right) M_0 ,
\end{equation}
with
\begin{equation}
r \equiv \sum_{i,j = 1}^2 \hat{\cal R}_{ij} = r_{11} + r_{22} - 2 r_{12}.
\end{equation}

In the case when $r_{ij}$ are much smaller compared to $\epsilon_{\nu i j}$ 
from Eqs.~(\ref{eq:see-saw}) and~(\ref{eq:Mr-1}) we get the left-handed neutrino Majorana mass 
matrix in the form:
\begin{equation}
M_{\nu_L} = - \frac{\lambda^2_\nu v^2_\nu}{4 M_{eff}} 
 \left[
     \left(\begin{array}{cc} 
        \epsilon^{\prime 2}_\nu  & \epsilon^\prime_\nu \epsilon_\nu \\ 
        \epsilon^\prime_\nu \epsilon_\nu & \epsilon^2_\nu \\
     \end{array} \right)  
   + r {\cal I} + O(r_{ij} \epsilon_{\nu ij} )
 \right] ,
\label{eq:MnuL}
\end{equation} 
where $\epsilon^\prime_\nu = \epsilon_{\nu 11} - \epsilon_{\nu 12}$ and  
$\epsilon_\nu = \epsilon_{\nu 12} - \epsilon_{\nu 22}$. 
The special form of ${\cal I}$ causes that $({\cal I} - {\cal E}_\nu ) \hat{\cal I} ({\cal I} - 
{\cal E}_\nu )^T = {\cal E}_\nu  \hat{\cal I} {\cal E}_\nu^T $ 
which corresponds to the first matrix in Eq.~(\ref{eq:MnuL}). 
The second term in Eq.~(\ref{eq:MnuL}) comes from ${\cal I} \hat{\cal R} {\cal I}$ and the last term includes
$- {\cal E}_\nu \hat{\cal R} {\cal I} - {\cal I} \hat{\cal R} {\cal E}_\nu^T + {\cal E}_\nu \hat{\cal R} {\cal 
E}_\nu^T$.
The eigenvalues of the first matrix in Eq.~(\ref{eq:MnuL}) are $\{ 0,\epsilon^{\prime 2}_\nu + 
\epsilon^2_\nu \}$ and 
the diagonalization matrix is given as:
\begin{equation}
U_{\nu_L} = 
     \left(\begin{array}{rr}
         - \frac{x}{\sqrt{1+x^2}} & \frac{1}{\sqrt{1+x^2}} \\
         \frac{1}{\sqrt{1+x^2}} & \frac{x}{\sqrt{1+x^2}} \\
     \end{array} \right)  , \quad x = \frac{\epsilon_\nu}{\epsilon^\prime_\nu}.
\end{equation}
In the limit $x \rightarrow \pm \infty $ or $x \rightarrow 0 $ this matrix 
is either diagonal or off-diagonal. Assuming further that the charged lepton diagonalization matrix has 
approximately the form of $U_{\cal I}$ in Eq.~(\ref{eq:UI}) we find that the lepton mixing matrix~(\ref{eq:U}) 
is approximately equal to $U_{\cal I}$ up to sign changes in different elements corresponding to different 
limits.
%
%
These situations lead to maximal mixing in the lepton sector and they occur
when one of $\epsilon_{11}$ and $\epsilon_{22}$ either dominates or 
is close to $\epsilon_{12}$. Both situations are 
reasonable to assume. 
No particular fine tuning is necessary. Close to 
maximal mixing, $U_{12} \sim 1/\sqrt{2} \pm 0.1 $, will be achieved every time when $|x| 
\gtrsim 7$ or $|x| \lesssim 1/7$.

The second term in Eq.~(\ref{eq:MnuL}) lifts  the first eigenvalue. In situations when large mixing in the 
lepton sector is generated, the masses of left-handed 
neutrinos are given as $\lambda^2_\nu v^2_\nu /(2 r M_0) \times \{r, \epsilon^2 \}$, 
where 
$\epsilon^2 =
max(\epsilon_\nu^2, \epsilon_\nu^{\prime 2})$ and
we assume $r < \epsilon^2$ and 
$r \gg  r_{ij}^2$ (in other words there is no cancellation between $r_{ij}$ which would make  $r$ 
smaller than higher order terms~\footnote{Such a cancellation is of course possible but unless it comes 
out naturally from some models it would require fine-tuning. We do not consider this possibility 
here. On the other hand, we do not have any reason to assume that $r$ is smaller than the 
dominant $\epsilon^2$. However, if this condition is not satisfied and the $r {\cal I}$ term in 
Eq.~(\ref{eq:MnuL}) dominates, the neutrino mass matrix resembles the charged lepton mass matrix and large 
hierarchy in neutrino masses and small mixing in the lepton sector would be generated.}). An interesting 
consequence of this approach is a very 
robust prediction for the mass of the lightest 
neutrino:
\begin{equation}
m_{\nu_1} = \lambda^2_\nu v^2_\nu /(2 M_0), 
\end{equation}
which is given 
by the 
universal Yukawa coupling $\lambda_\nu$ and the
overall right-handed neutrino mass scale $M_0$. It does not depend 
either on details of the Yukawa matrix ($\epsilon_{\nu ij}$) or details of the right-handed neutrino 
Majorana mass matrix ($r_{ij}$). We discuss this more in the case of three families. 

Before we discuss three families, let us summarize why it was possible to obtain large mixing 
in the lepton sector. The form of $\hat{\cal I}$ given in Eq.~(\ref{eq:hatIR}) plays the 
crucial role.
The sum of elements of this matrix in every row and column is zero. As a
result, ${\cal I} \hat{\cal I} = 0$ and so 
the 1s from the neutrino Yukawa matrix wash out, leaving products of $\epsilon_{\nu 
ij}$ as the 
dominant contributions to the left-handed neutrino Majorana mass matrix after the seesaw.
If the resulting $M_{\nu_L}$ is hierarchical (and it is if we assume that ${\cal E}_\nu$ is similar to 
the perturbation matrices for other fermions), 
the lepton mixing matrix will be dominated by $U_e$. Perhaps the only 
nontrivial assumption is that $r < \epsilon^2$. However, this assumption does not require a departure 
from a democratic approach. Quite the contrary, it just means that $M_{\nu_R}$ has to be somewhat more 
democratic than Dirac Yukawa matrices.  
We will see that the matrix $\hat{\cal I}$ has the same property 
in the case of three families although the form 
of this matrix and the resulting left-handed neutrino Majorana mass matrix 
is much more complicated.

\section{\label{sec:3g} Democratic matrices: three families}

Let us now assume that all three generations are indistinguishable in leading 
order and so Yukawa couplings are given as:  
\begin{equation}
Y_f \equiv \frac{1}{3} \lambda_f \left( {\cal I} - {\cal E}_f \right) , \quad 
{\cal E}_f = \left(\begin{array}{ccc} \epsilon_{f 11} & \epsilon_{f 12} & \epsilon_{f 13} \\
                                      \epsilon_{f 12} & \epsilon_{f 22} & \epsilon_{f 23} \\
                                      \epsilon_{f 13} & \epsilon_{f 23} & \epsilon_{f 33}
                   \end{array} \right) ,
\end{equation}
where
we use the same symbol ${\cal I}$ for the $3 \times 3$ matrix with unit elements as we did in 
the $2 \times 2$ case, and similarly we  parametrize the departure from universality by 
matrices ${\cal E}_f$.
If  Yukawa matrices were equal to ${\cal I} \lambda_f /3 $, then mass eigenvalues are
$\{ 0,0,\lambda_f \}$ and the diagonalization matrix is:
\begin{equation}
U_{\cal I} = 
\left(\begin{array}{ccc}
        \ \  \cos \theta_{\cal I}   & \sin \theta_{\cal I}     & \ \ 0 \\
        - \sin \theta_{\cal I} & \cos \theta_{\cal I}     & \ \ 0 \\
       \ \   0               & 0                 & \ \ 1
   \end{array} \right) \,
            \left(\begin{array}{rrc}  
      \frac{1}{\sqrt{2}} & -\frac{1}{\sqrt{2}} &  \ \ \ 0 \\
      \frac{1}{\sqrt{6}} &  \frac{1}{\sqrt{6}} & -\frac{2}{\sqrt{6}}\\
      \frac{1}{\sqrt{3}} &  \frac{1}{\sqrt{3}} &  \ \ \frac{1}{\sqrt{3}}
                 \end{array} \right)  .
\label{eq:UI3}
\end{equation}
As a consequence of  degenerate zero eigenvalues the first two rows of this matrix are not uniquely 
specified and are model dependent ($\cal E$ has to be taken into account). They can be replaced by 
any of their linear combinations and the corresponding orthogonal combination, which is accounted for by 
the first matrix which rotates the first two rows. 
As a result, the CKM matrix is not the identity 
matrix in the leading order as it was in the case of two families, but rather a unitary matrix 
with an arbitrary 
1-2 element~\footnote{Therefore the Cabibbo angle is not necessarily related to the hierarchy in quark 
masses. Other off-diagonal elements, 1-3 and 
2-3, are zero in the leading order and their exact values are  related to the generated 
hierarchy in quark sector. 
}.

The Majorana mass matrix for right-handed neutrinos is parametrized in a similar way as 
before:
\begin{equation}
M_{\nu_R} = \frac{1}{3} \left( {\cal I} - {\cal R} \right) M_0 , \quad
{\cal R} = \left(\begin{array}{ccc}  r_{11} & r_{12} & r_{13} \\
                                     r_{12} & r_{22} & r_{23} \\
                                     r_{13} & r_{23} & r_{33} \\
                 \end{array} \right) .
\end{equation}
The inverse of this matrix is given as:
\begin{equation}
M_{\nu_R}^{-1} = \frac{1}{M_{eff}} \left( \hat{\cal I} + \hat{\cal R} \right),
\end{equation}
where $\hat{\cal I}$ can be written as
\begin{equation}
\hat{\cal I} \equiv 
   \left(\begin{array}{ccc}
      - r_1 & r_1 - r_3            & + r_3 \\
  r_1 - r_3 & - r_1 + 2 r_3 - r_2  & - r_3 + r_2 \\
      + r_3 & - r_3 + r_2            & - r_2  
   \end{array} \right) ,
\label{eq:hatI3} 
\end{equation}
with
\begin{eqnarray}
r_1 &=& r_{22} + r_{33} - 2 r_{23} ,\\
r_2 &=& r_{11} + r_{22} - 2 r_{12} ,\\
r_3 &=& r_{22} + r_{13} - r_{12} - r_{23} .
\end{eqnarray}
From the form of $\hat{\cal I}$ it is easy to see the correspondence with the $2 \times 2$ 
case, since it can also be written as:
\begin{equation}
\hat{\cal I} = r_1 \hat{\cal I}_1 + r_2 \hat{\cal I}_2 + r_3 \left( \hat{\cal I}_3
+ \hat{\cal I}_3^T \right) ,
\label{eq:invI2}
\end{equation}
where
\begin{equation}
\hat{\cal I}_1 \equiv 
    \left(\begin{array}{rrr}
      - 1 & 1   & 0 \\
        1 & - 1 & 0 \\
        0 & 0   & 0
   \end{array} \right) , \quad
\hat{\cal I}_2 \equiv
   \left(\begin{array}{rrr}
        0 & 0   & 0 \\
        0 & - 1 & 1 \\
        0 & 1   & -1
   \end{array} \right) , \quad
\hat{\cal I}_3 \equiv
\left(\begin{array}{rrr}
        0 & - 1   & 1 \\
        0 &   1 &   -1 \\
        0 & 0   & 0
   \end{array} \right).
\end{equation}
The matrix $\hat{\cal R}$ is proportional to the inverse of the matrix $\cal R$.
Its elements are cofactors of the corresponding elements of $\cal R$:
\begin{equation}
\hat{\cal R}_{ij} = \frac{1}{2} \epsilon_{ikl} \epsilon_{jmn} r_{km} r_{ln}.
\end{equation}
And finally, $M_{eff}$ is given as
\begin{equation}
M_{eff} = \frac{1}{3} \left( r - \det{\cal R} \right) M_0 ,
\label{eq:Meff3}
\end{equation}
where
\begin{equation}
r \, \equiv \, \sum_{i,j = 1}^3 \hat{\cal R}_{ij} \, = \, r_1 r_2 - r_3^2 .
\end{equation}
Let us again assume that $r_{ij}$ are much smaller than $\epsilon_{\nu i j}$. In this 
case we get:
\begin{equation}
M_{\nu_L} = - \frac{\lambda^2_\nu v^2_\nu}{9 M_{eff}}
 \left[ {\cal M} + r {\cal I} + O(\hat{\cal R}_{ij} \epsilon_{\nu ij} )
 \right] ,
\label{eq:MnuL3}
\end{equation}
where ${\cal M} \equiv {\cal E}_\nu \hat{\cal I} {\cal E}_\nu^T $ is a $3 \times 3$ equivalent to the 
first matrix in Eq.~(\ref{eq:MnuL}). 
As in the case of two families the matrix $\cal M$ depends on differences between $\epsilon_{\nu ij}$.
It is useful to define the perturbation vectors:
\begin{eqnarray}
\vec e & = & {\cal E}_{\nu 1} - {\cal E}_{\nu 2} , \\
\vec g & = & {\cal E}_{\nu 2} - {\cal E}_{\nu 3} ,
\end{eqnarray} 
where ${\cal E}_{\nu i}$ is the i-th column of the perturbation matrix ${\cal E}_\nu$. 
The matrix $\cal M$ can be written as:
\begin{equation}
{\cal M} = - \, r_1 \, \left( \vec e \, . \, {\vec e}^{\; T} \right) 
           \; - \; r_2 \, \left( \vec g \, . \, {\vec g}^{\; T} \right)
           \; - \; r_3 \, \left( \vec e \, . \, {\vec g}^{\; T} + \vec g \, . \, {\vec e}^{\; T} 
                 \right) .
\end{equation}
From this expression it is easy to see that ${\cal M} \, {\vec v}_0 = 0 $ for ${\vec v}_0 \perp {\vec e}$, 
${\vec v}_0 \perp {\vec g}$. Therefore, the eigenvector corresponding to the zero eigenvalue is given as:
\begin{equation}
{\vec v}_0 = \frac{{\vec e} \times {\vec g}}{|{\vec e} \times {\vec g} |}.
\end{equation} 
The heavy two eigenvalues can be written as:
\begin{equation}
m_\pm = \frac{1}{2} t \left( 1 \pm \sqrt{1 - \frac{4d}{t^2} } \right),
\label{eq:mpm}
\end{equation}
with
\begin{eqnarray}
t & = & \rho_1 + \rho_2 + 2\rho_3 \cos \alpha , \\
d & = & \left( \rho_1 \rho_2 - \rho_3^2 \right) \sin^2 \alpha ,               
\end{eqnarray}
where
\begin{equation}
\rho_1 \equiv - r_1 \, | \, \vec e \, |^2, \quad 
\rho_2 \equiv - r_2 \, | \, \vec g \,  |^2, \quad 
\rho_3 \equiv - r_3 \, | \, \vec e \,  | \, | \, \vec g \, |,
\label{eq:rho}
\end{equation}
and $\alpha$ is the angle between ${\vec e}$ and ${\vec g}$.
The eigenvectors corresponding to these two eigenstates are given as two orthogonal linear 
combinations of ${\vec e}$ and ${\vec g}$:
\begin{equation}
{\vec v}_\pm = a_\pm \, {\vec e} \; + \; b_\pm \, {\vec g},
\end{equation}
where $a_\pm, \, b_\pm$ can be written in terms of $\rho_1$, $\rho_2$, $\rho_3$, and $\cos 
\alpha$.
Before we proceed further let us summarize the current status of neutrino masses and mixing.

\subsection{Experimental results}

A global analysis of neutrino oscillation data~\cite{Gonzalez-Garcia:2003, Maltoni:2003}
gives the best fit to the neutrino mass-squared differences: 
\begin{eqnarray}
\Delta m^2_{sol} & \equiv m_{\nu_2}^2 - m_{\nu_1}^2 & \simeq  6.9 \times 10^{-5} {\rm eV}^2, \\
\Delta m^2_{atm} & \equiv m_{\nu_3}^2 - m_{\nu_1}^2 & \simeq  2.6 \times 10^{-3} {\rm eV}^2,
\end{eqnarray}
and mixing angles:
\begin{eqnarray}
\sin^2 \theta_{12} & \equiv & \sin^2 \theta_{sol} \;  = \;  0.30 \, , \label{eq:sol_c} \\
\sin^2 \theta_{23} & \equiv & \sin^2 \theta_{atm} \;  = \;  0.52 \, . \label{eq:atm_c}  
\end{eqnarray}
The $3 \sigma$ ranges for mixing angles are:
\begin{eqnarray}
0.23 \leq & \sin^2 \theta_{sol} & \leq 0.39 , \label{eq:sol_3s} \\
0.31 \leq & \sin^2 \theta_{atm} & \leq 0.72 , \label{eq:atm_3s}
\end{eqnarray} 
and the $3 \sigma$ upper bound on the third mixing angle is:
\begin{equation}
\sin^2 \theta_{13} \; \leq \; 0.054 \, \label{eq:13_3s} . 
\end{equation} 
In the case $m_{\nu_1} \ll m_{\nu_2}, \, m_{\nu_3}$ we can interpret these results as: 
\begin{eqnarray}
m_{\nu_2} & \simeq & \sqrt{\Delta m^2_{sol}} \; \simeq  8.3 \times 10^{-3} {\rm eV}, \label{eq:mnu2} \\
m_{\nu_3} & \simeq & \sqrt{\Delta m^2_{atm}} \; \simeq  5.1 \times 10^{-2} {\rm eV}, \label{eq:mnu3}
\end{eqnarray}
and due to $\sin^2 \theta_{13} \simeq 0$, the mixing angles are related to the elements of
the lepton mixing matrix in the following simple way:
\begin{eqnarray}  
\sin^2 \theta_{12} & \simeq & |\, U_{12} \, |^2 \, , \\
\sin^2 \theta_{23} & \simeq & |\, U_{23} \, |^2 \, .
\end{eqnarray}

\subsection{Back to three families} 

From the experimental results above we see that the lepton mixing matrix is characterized by a very small 1-3 
mixing and close to maximal 2-3 mixing. The lepton mixing matrix originates from both the charged lepton 
diagonalization 
matrix $U_e$ and the neutrino diagonalization matrix $U_{\nu_L}$. In a democratic approach the charged lepton 
diagonalization matrix already contains large mixing angles. If also $U_{\nu_L}$ contains large mixing 
angles, it would require a conspiracy between elements of all perturbation matrices in order to achieve $\sin^2 
\theta_{13} \simeq 0$ and $\sin^2 \theta_{23} \simeq 0.5$. Of course, such a conspiracy might naturally 
occur in some models. 

\begin{figure}
\includegraphics{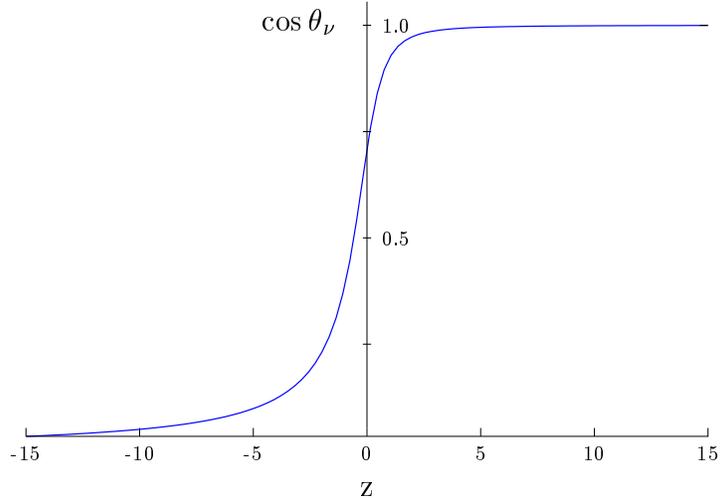} 
\caption{\label{fig:cos_th_n} $\cos \theta_\nu$ as a function of $z$, see Eq.~(\ref{eq:cos_th_n}).}
\end{figure}

In order to avoid any exact relations between elements of ${\cal E}_e$, ${\cal E}_\nu$, and ${\cal R}$   
the simplest way to proceed is to assume that the perturbation matrix ${\cal E}_\nu$ introduces the minimal 
amount of mixing into the lepton mixing matrix. This happens when perturbation vectors $\vec e$ and $\vec 
g$ are dominated by a single element $\sim \, (0,0,1)^T$, $(0,1,0)^T$, or $(1,0,0)^T$ with $\vec e \perp \vec g$.
In this case $U_{\nu_L}$ is given as:
\begin{equation}
U_{\nu_L} =
    \left(\begin{array}{ccc}
        1 & 0   & 0 \\
        0 & \cos \theta_\nu   & - \sin \theta_\nu \\
        0 & \sin \theta_\nu   & \cos \theta_\nu 
   \end{array} \right) P 
   \left(\begin{array}{ccc}
        1 & 0  & 0 \\
        0 & 1  & 0 \\
        0 & 0  & 1
   \end{array} \right),
\label{eq:UnuL}
\end{equation} 
where $P$ is a permutation matrix which interchanges rows of the identity matrix depending on the choice of 
perturbation vectors $\vec e$ and $\vec g$. Note that, since $\cos \alpha = 0$, the eigenvalues and $\cos 
\theta_\nu$ are given in terms of 3 parameters: $\rho_1$, $\rho_2$ and $\rho_3$. Namely, in 
Eq.~(\ref{eq:mpm}) $t = \rho_1 + \rho_2 $ and $d = \rho_1 \rho_2 - \rho_3^2$, and in
Eq.~(\ref{eq:UnuL}),
\begin{equation}
\cos \theta_\nu = \frac{1}{\sqrt{2 + 2 z^2 - 2z \sqrt{1+z^2}}}
\label{eq:cos_th_n}
\end{equation}
(up to an overall sign), where
\begin{equation}
z \equiv \frac{\rho_2 - \rho_1}{2 \rho_3}.
\label{eq:z}
\end{equation}
This function is plotted in Fig.~\ref{fig:cos_th_n}.
We see that $\cos \theta_\nu$ varies fast for small $| z |$ and is almost constant for $| z | \gg 0$. 
Therefore, the least-fine-tuned situations correspond to $| z | \gg 0$ which happens for 
$| \rho_3 | \ll \max \{|\rho_1 |, | \rho_2 | \}$. Actually, no strong hierarchy is necessary, $\cos \theta_\nu$ 
is 
very close to 1 for $z$ as small as 3. Note that, if we assume $| \, \vec e \,  | \, \ll \, | \, \vec g \, |$, which is 
reasonable to assume about all perturbation matrices in order to generate hierarchy between generations,
$| \rho_3 | \ll \max \{|\rho_1 |, | \rho_2 | \}$ is satisfied for a huge variety 
of possible  entries $r_{ij}$.

From Eqs.~(\ref{eq:U}), (\ref{eq:UI3}) and (\ref{eq:UnuL}) we see that the most general form of the lepton 
mixing 
matrix in this case can be written as~\footnote{In the most general case there are additional three angles 
in this matrix corresponding to the overall rotation of the given perturbation vectors ${\vec e}$ and 
${\vec g}$ to 
the basis chosen here. One of those angles and also $\cos \alpha$ can be always absorbed into $\cos 
\theta_\nu$.}:
\begin{equation}
U = \left(\begin{array}{ccc}
        \cos \theta_e   & \sin \theta_e     & 0 \\
        - \sin \theta_e & \cos \theta_e     & 0 \\
        0               & 0                 & 1
   \end{array} \right)
\left(\begin{array}{rrc}
      \frac{1}{\sqrt{2}} & -\frac{1}{\sqrt{2}} &  \ \ \ 0 \\
      \frac{1}{\sqrt{6}} &  \frac{1}{\sqrt{6}} & -\frac{2}{\sqrt{6}}\\
      \frac{1}{\sqrt{3}} &  \frac{1}{\sqrt{3}} &  \ \ \frac{1}{\sqrt{3}}
                 \end{array} \right)
   \left(\begin{array}{ccc}
        1 & 0   & 0 \\
        0 & \cos \theta_\nu     & \sin \theta_\nu \\
        0 & - \sin \theta_\nu   & \cos \theta_\nu
   \end{array} \right) ,
\label{eq:U3}
\end{equation}
where the permutation matrix from Eq.~(\ref{eq:UnuL}) was absorbed in the redefinition of $\cos \theta_e$ since 
the permutation matrix switches columns of the second matrix above and this can be accounted for by rotation of 
the first two rows of that matrix.

\begin{figure}
\includegraphics{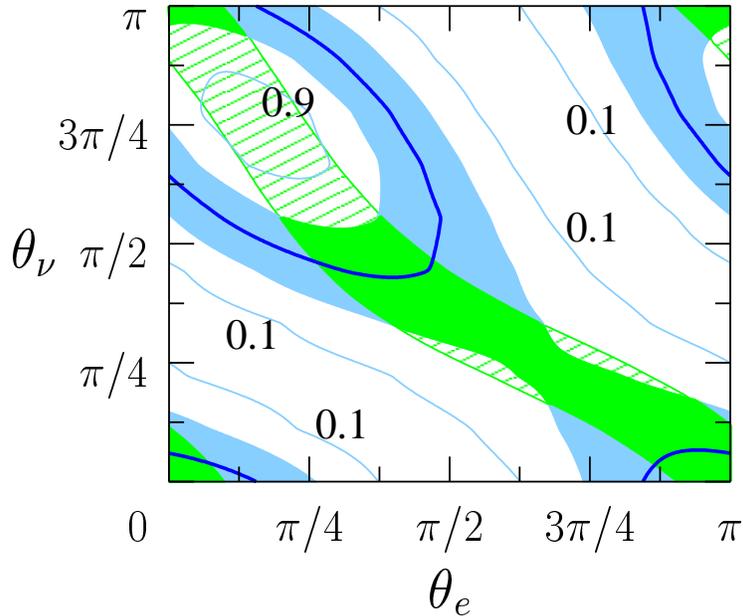}
\caption{\label{fig:atm} Contours of constant $\sin^2 \theta_{23}$ (blue) as a function of
$\theta_e$ and $\theta_\nu$. Dark blue contours represent the central value and
the blue shaded area corresponds to the $3 \sigma$ allowed region. Overlayed is the $3 \sigma$ region of
$\sin^2 \theta_{13}$ (green stripes). The green shaded area represents the overlap of $3 \sigma$
allowed regions of $\sin^2 \theta_{23}$ and $\sin^2 \theta_{13}$.} \end{figure}
The lepton mixing matrix in this simplest scheme is given in terms of two parameters $\theta_e$ and 
$\theta_\nu$, and so it is not trivial that the resulting three mixing angles can be simultaneously within 
experimental bounds, although, from the suggestive form of the matrix in the middle of Eq.~(\ref{eq:U3}), it 
might 
be guessed that it will happen for small $\theta_e$ and $\theta_\nu$. 
In Fig.~\ref{fig:atm} we present contours of constant $\sin^2 \theta_{23}$ (blue) as a function of $\theta_e$ 
and $\theta_\nu$. Dark blue contours represent the central value (Eq.~(\ref{eq:atm_c})) and
the blue shaded area corresponds to the $3 \sigma$ allowed region (Eq.~(\ref{eq:atm_3s})). Green stripes 
represent the $3 \sigma$ allowed region of $\sin^2 \theta_{13}$ given in Eq.~(\ref{eq:13_3s}) and the green 
area is the overlap of $3 \sigma$ allowed regions of $\sin^2 \theta_{23}$ and $\sin^2 \theta_{13}$.
We see that it is not particularly difficult to achieve a very small $\theta_{13}$ and close to maximal 
$\theta_{23}$. In the majority of cases, the value of $\sin^2 \theta_{23}$ consistent with the limits on $\sin^2 
\theta_{13}$ is within or above the  $3 \sigma$ allowed region.

\begin{figure}
\includegraphics{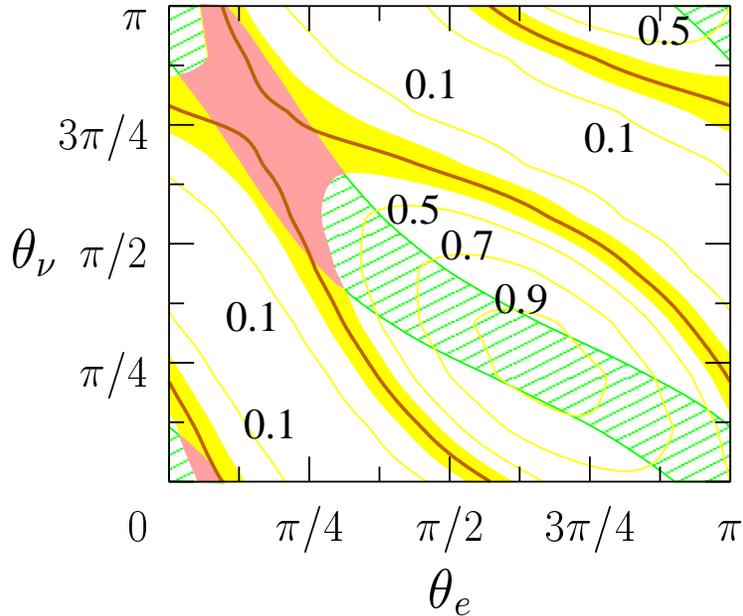}
\caption{\label{fig:sol} Contours of constant $\sin^2 \theta_{12}$ (yellow) as a function of
$\theta_e$ and $\theta_\nu$. Brown contours represent the central value and
the yellow shaded area corresponds to the $3 \sigma$ allowed region. Overlayed is the $3 \sigma$ region
of
$\sin^2 \theta_{13}$ (green stripes). The pink shaded area represents the overlap of $3 \sigma$
allowed regions of $\sin^2 \theta_{12}$ and $\sin^2 \theta_{13}$.}
\end{figure}
In Fig.~\ref{fig:sol} we present a similar plot for the solar mixing angle. Contours of constant $\sin^2 
\theta_{12}$ are represented by yellow. Brown contours correspond to the central 
value (Eq.~(\ref{eq:sol_c})) and
the yellow shaded area represents the $3 \sigma$ allowed region (Eq.~(\ref{eq:sol_3s})). 
Green stripes have the same meaning as in the Fig.~\ref{fig:atm} and the overlap of $3 \sigma$ allowed 
regions of $\sin^2 \theta_{12}$ and $\sin^2 \theta_{13}$ is given by the pink area.
The value of $\sin^2 \theta_{12}$ consistent with the limits on $\sin^2
\theta_{13}$ is always within or above the  $3 \sigma$ allowed region.

\begin{figure}
\includegraphics{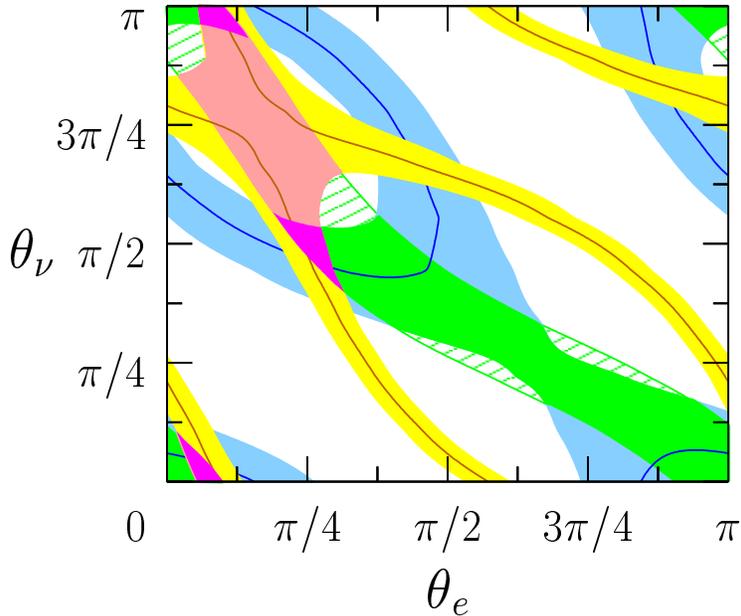}
\caption{\label{fig:all} The $3 \sigma$ allowed regions of $\sin^2 \theta_{23}$ (blue, green)),
$\sin^2 \theta_{12}$ (yellow, pink) and $\sin^2 \theta_{13}$ (green, pink) from Fig.~\ref{fig:atm}
and Fig.~\ref{fig:sol} overlayed. The magenta regions represent their overlap. }
\end{figure}

However, the $3 \sigma$ allowed regions of all three mixing angles overlap only in two small regions 
as can be seen in Fig.~\ref{fig:all}. 
These regions are represented by magenta (note the periodicity of the picture; regions disconnected at the 
boundaries of the plot are not counted separately).

\subsection{How small can $\sin^2 \theta_{13}$ be?}

Since we do not measure $\theta_{13}$, 
it is interesting to ask what values of $\sin^2 \theta_{13}$ can be achieved within this
approach while satisfying the $3 \sigma$ experimental bounds of $\sin^2 \theta_{23}$ and $\sin^2 \theta_{12}$.
In Fig.~\ref{fig:all} we see that the $3 \sigma$ regions of $\sin^2 \theta_{23}$ and $\sin^2 \theta_{12}$ 
overlap in 4 places: two regions containing magenta areas (Region I), and the regions around points
$\theta_e \simeq \pi/2$, $\theta_\nu \simeq 5 \pi/8$ and  $\theta_e \simeq 7\pi/8$, $\theta_\nu \simeq 7\pi/8$ 
(Region II). Note that the two areas in the Region I predict $\sin
\theta_{13}$ with opposite signs, and the same applies to the Region II.
The predicted values of $\sin^2 \theta_{13}$ from these regions are: 
\begin{eqnarray}
0.008 \leq & \sin^2 \theta_{13} & \leq 0.14  \quad \quad {\rm (Region \; I)} , \\
0.22 \leq & \sin^2 \theta_{13} & \leq 0.66  \quad \quad {\rm (Region \; II)} .
\end{eqnarray}
Region I overlaps with the experimentally allowed region and it shows that the value of $\sin^2 
\theta_{13}$ which can be accommodated in this approach can be as low as 0.008.
Note that this minimal value of $\sin^2
\theta_{13}$ corresponds to the maximal allowed values of $\sin^2 \theta_{23}$ and 
$\sin^2 \theta_{12}$. On the other hand, the central values of $\sin^2 \theta_{23}$ and
$\sin^2 \theta_{12}$ correspond to $\sin^2 \theta_{13}$ near its present experimental upper bound.

\subsection{\label{sec:consequences} Consequences for models}

The magenta regions in Fig.~\ref{fig:all} are not large. Nevertheless, the good news is that they are 
located around $\theta_\nu \simeq \pi/2$ or $\theta_\nu \simeq 0, \pi$, which corresponds to the 
least-fine-tuned possibilities discussed after~Eq.~(\ref{eq:z}).    

The only allowed values of $\theta_e$ are close to $0$ or $\pi/4$. The charged lepton diagonalization matrix 
corresponding to these two possibilities is close to the forms:
\begin{equation}
U_e = \left(\begin{array}{rrc}
      \frac{1}{\sqrt{2}} & -\frac{1}{\sqrt{2}} &  \ \ \ 0 \\
      \frac{1}{\sqrt{6}} &  \frac{1}{\sqrt{6}} & -\frac{2}{\sqrt{6}}\\
      \frac{1}{\sqrt{3}} &  \frac{1}{\sqrt{3}} &  \ \ \frac{1}{\sqrt{3}}
                 \end{array} \right)
\label{eq:Ue1}
\end{equation}
for $\theta_e = 0$, and 
\begin{equation}
U_e = \frac{1}{2 \sqrt{3}} \, 
              \left(\begin{array}{ccc}
      -\sqrt{3} -1 & \ \ \sqrt{3} -1 & \ \  2 \\
      \ \ \sqrt{3} -1 & -\sqrt{3} -1 & \ \  2 \\
      \ \ 2 &  \ \ 2  & \ \  2
                 \end{array} \right)
\end{equation}
for $\theta_e = \pi/4$ (up to overall sign changes in the rows).
The first solution is obvious, since $U_e$ is very close to the observed lepton mixing matrix in this case.
The second solution is not so obvious and its symmetric form is quite surprising.  

Perhaps the simplest models of this kind are those with a dominant 3-3 component in all 
perturbation matrices, ${\cal E} \simeq {\cal R} \simeq {\rm diag} (0,0,\epsilon)$. This perturbation generates 
masses 
of the second 
family. In this case $U_e$ is very close to the one in~Eq.~(\ref{eq:Ue1}) and $\vec g \simeq (0,0,1)^T$. 
There are many ways to introduce masses of the first family. Simple examples are ${\cal E} \simeq {\cal R} \simeq 
{\rm diag} (0,\delta,\epsilon)$ (this perturbation was suggested in~\cite{branco_efd})
or ${\rm diag} (\delta ,0,\epsilon)$, where $\delta \ll \epsilon$ (with 
different numerical values for different perturbation matrices). These 
situations have all the desired features: $\vec e \perp \vec g$ (approximately) and $|z| \gg 0$.

It is certainly remarkable that very simple forms of perturbation matrices, which on top of everything can be 
chosen to be the same for all mass matrices, lead to the observed pattern 
of fermion masses and mixing. This should come with a warning, however. The simple form of perturbation matrices 
does not guarantee that it is easy to obtain them naturally in some models. Especially in models with family 
symmetries this is very complicated compared to the situation in hierarchical models. The reason is that in 
hierarchical models hierarchy is achieved by suppressing or forbidding some entries in mass 
matrices. This can be easily achieved by assigning different charges under a family symmetry to different 
particles. In democratic models however, family symmetries have to allow all entries in mass matrices and yet 
they have to 
account for small differences in specific elements of mass matrices. This is certainly not trivial to achieve.

\subsection{Mass of the lightest neutrino}

The mass of the lightest neutrino is lifted when the second term in Eq.~(\ref{eq:MnuL3}) is taken into
account. Since we assume it is just a small correction to the first term, it can be treated as a
perturbation. Adding this perturbation does not significantly affect the two heavy eigenvalues and
the diagonalization matrix, but it is crucial for the lightest eigenvalue which is exactly zero in
the limit when this term is ignored. 
In the case of non-degenerate eigenvalues the correction to eigenvalues $m_i$ of a matrix $\cal M$ generated 
by a matrix $\delta \cal M$ are given as:
\begin{equation}
\delta m_i = u_i^T \, \delta {\cal M} \, u_i , 
\end{equation}
where $u_i$ are normalized eigenvectors. In our case $\delta {\cal M} = r {\cal I}$ (up to the overall factor 
in Eq.~(\ref{eq:MnuL3})) and the eigenvector corresponding to the zero eigenvalue is ${\vec v}_0 \simeq (1,0,0)^T$.
Therefore,
\begin{equation}
m_{\nu_1} =  \frac{\lambda^2_\nu v^2_\nu}{9 M_{eff}} r .
\end{equation}
Since $M_{eff} \simeq r M_0/3$, see Eq.~(\ref{eq:Meff3}), the mass of the lightest neutrino again does not depend 
on details of a model in the leading order and is given as:
\begin{equation}
m_{\nu_1} = \frac{\lambda_\nu^2 v_\nu^2}{3M_0}.
\label{eq:m_nu1a}
\end{equation}
This result is based on our assumption of  the minimal amount of mixing coming from the neutrino diagonalization 
matrix. However, it is possible to make a prediction which does not depend on this assumption.

Let us suppose that we do not know what the eigenvector corresponding to the lightest eigenvalue is.
Due to the universal form of $\delta {\cal M}$, we have
\begin{equation}
{\vec v}_0^{\, T} \, \delta {\cal M} \, {\vec v}_0 = r \, \xi^2 , 
\label{eq:pert}
\end{equation}
where
\begin{equation}
\xi = \sum_{i=1}^3 v_{0i} ,
\end{equation}
and so the mass of the lightest neutrino is given as:
\begin{equation}
m_{\nu_1} =  \frac{\lambda^2_\nu v^2_\nu}{3M_0} \, \xi^2 .
\label{eq:m_nu1}
\end{equation}
In general $\xi$ can be anything between 0 and $\sqrt3$. However, in order to satisfy bounds on lepton mixing 
angles $\xi$ cannot be arbitrary. The 3-1 element of the lepton mixing matrix is given by:
\begin{equation}
U_{\tau 1} = \left( U_e U_{\nu_L}^\dag \right)_{31} = \left( \frac{1}{\sqrt{3}}, \frac{1}{\sqrt{3}}, 
\frac{1}{\sqrt{3}} \right) . {\vec v}_0 = \frac{1}{\sqrt{3}} \xi ,
\label{eq:U31_xi}
\end{equation}
and so 
\begin{equation}
m_{\nu_1} =  \frac{\lambda^2_\nu v^2_\nu}{M_0} \, U_{\tau 1}^2 .
\label{eq:m_nu1b}
\end{equation}
Note that the 3rd row in $U_e$ {\it is not} model dependent unlike the first two rows are! It can receive 
only small 
corrections from the perturbation matrix. Finally, in the case of complex matrices, ${\vec v}^{\, T}$ in Eq.~(\ref{eq:pert}) 
becomes ${\vec v}^{\, \dag}$ and $U_{\tau 1}^2$ in Eq.~(\ref{eq:m_nu1b}) becomes  $| \, U_{\tau 1} \, |^2$.

Although we do not measure $U_{\tau 1}$, it is related to the observed mixing angles due to the unitarity of the 
lepton mixing matrix. In the case $\sin \theta_{13} \simeq 0$ it is simply given by
\begin{equation}
U_{\tau 1} = \sin \theta_{23} \, \sin \theta_{12} .
\end{equation}
A global analysis of neutrino oscillation data~\cite{Gonzalez-Garcia:2003}
gives the $3 \sigma$ range:
\begin{equation}
0.20 \leq \, | \, U_{\tau 1} \, | \, \leq 0.58 .
\label{eq:U31_exp}
\end{equation}
The value of $U_{\tau 1} = 1/\sqrt{3}$ in which case Eq.~(\ref{eq:m_nu1b}) gives the same result as 
Eq.~(\ref{eq:m_nu1a}) is close to the 
upper limit.

The masses of the two heavier neutrinos are given in terms of $r_{ij}, \, \epsilon_{\nu kl}$, and so they 
are
 highly model dependent. Let us look at a simple example to get a feeling for  typical values of perturbations
which lead to the observed spectrum. Let us assume that the form of all perturbations is $\sim {\rm diag} (\delta
,0,\epsilon)$. In this case we have: $| \, \vec e \, |^2 = \delta_\nu^2$, $ |\,  \vec g \,  |^2 =
\epsilon_\nu^2$, $ r_1 = \epsilon_r$, $ r_2 = \delta_r$, and $ r_3 = 0$, from which we get: $\rho_1 = \epsilon_r  
\delta_\nu^2$,
$\rho_2 = \delta_r  \epsilon_\nu^2$, $\rho_3 = 0$, and $r = r_1 r_2$. The neutrino masses are given by:
\begin{equation}
m_{\nu_{2,3}} = \frac{\lambda^2_\nu v^2_\nu}{9 M_{eff}} \, \left\{ \rho_1, \rho_2 \right\} 
\simeq \frac{\lambda^2_\nu v^2_\nu}{3 M_0 \ r} \, \left\{ \rho_1, \rho_2 \right\}
\simeq \frac{\lambda^2_\nu v^2_\nu}{3 
M_0 } \, \left\{ \frac{\delta_\nu^2}{ \delta_r}, \frac{\epsilon_\nu^2}{ \epsilon_r} \right\}.
\end{equation}
Using Eq.~({\ref{eq:m_nu1b}) we get
\begin{equation}
m_{\nu_{2,3}} \simeq \frac{m_{\nu_1}}{3 \, | \, U_{\tau 1} \, |^2}  \, \left\{ \frac{\delta_\nu^2}{ \delta_r}, 
\frac{\epsilon_\nu^2}{ \epsilon_r} \right\}.
\label{eq:mnu23}
\end{equation}
In order to have $m_{\nu_2}, m_{\nu_3} > m_{\nu_1}$ we need $\delta_r < \delta_\nu^2$ and $\epsilon_r < 
\epsilon_\nu^2$, 
and 
so the hierarchy in the right-handed neutrino mass matrix has to be much larger than the hierarchy in the neutrino Yukawa 
matrix.  This coincides with the assumption we had to make in order to achieve large lepton mixing, see 
Eq.~(\ref{eq:MnuL3}).

In
simple SO(10) type models $\lambda_u v_u = \lambda_\nu v_\nu$, in which case the lightest and the
heaviest fermion of the standard model
are connected through the relation in Eq.~(\ref{eq:m_nu1b}) where $\lambda_\nu^2 v_\nu^2$ is replaced by  
$m_{top}^2$ (actually, to be precise, $\lambda_u = \lambda_\nu$ is a relation at the GUT scale and the effects of
the renormalization group running between the GUT scale and the electroweak scale should be taken into account). 
This is a very pleasant feature since
we can further 
identify $M_0$ with the GUT scale, $M_{GUT} \sim 2 \times 10^{16}$ GeV, in which case we get
\begin{equation}
m_{\nu_1} = \frac{m_{top}^2}{M_{GUT}} \, | \, U_{\tau 1} \, |^2 ,
\label{eq:m_nu1c}
\end{equation}
and predict the mass of the lightest neutrino
to be between $ 5 \times 10^{-5}$ eV and  $ 5 \times 10^{-4}$ eV depending on the value of $U_{\tau 1}$. 

From experimental values of 
$m_{\nu_2}$ and $m_{\nu_3}$, given in Eqs.~(\ref{eq:mnu2}) and (\ref{eq:mnu3}), and from the predicted mass of
$m_{\nu_1}$, we see that $\epsilon_\nu^2 / \epsilon_r $ in Eq.~(\ref{eq:mnu23}) is of order $10^2$
and $\delta_\nu^2 / \delta_r $ is an order of magnitude smaller. The largest $\epsilon_f$ necessary to fit the  
masses of the second generation of charged 
fermions corresponds to $\epsilon_e \simeq 0.28$. If we take $\epsilon_\nu \simeq 0.1$, we find $\epsilon_r \simeq 
10^{-4} $ and the mass of the second right-handed neutrino is $M_2 \simeq (10^{-4} - 10^{-5}) M_3$, 
where $M_3 = M_0 = M_{GUT}$. This is a very rough estimate and it is not clear 
how to reasonably estimate $M_1$, besides the relation $M_1 < M_2$. We conclude that the spectrum of the right-handed 
neutrinos consistent with bi-large mixing in our setup is $M_1 < M_2 < 10^{-4} M_3$ and  $M_3 \simeq M_{GUT}$.

\section{\label{sec:symmetries} A Symmetry for Third Generation Dominance}

The democratic Yukawa matrices are well motivated by $S_{3L} \times S_{3R}$ family
symmetry~\cite{Fritzsch_Xing_review}. However, the right-handed neutrino Majorana mass matrix is then
constrained by the $S_{3R}$ symmetry only and the democratic form is not unique.
The most general form of the right-handed neutrino mass matrix is given as a linear combination
of the democratic matrix and the identity matrix.

It has been shown that the democratic forms of Yukawa matrices and the right-handed neutrino mass matrix
are uniquely specified by imposing
a $Z_3$ symmetry realized in the following way~\cite{branco_Z3}:
\begin{eqnarray}
f_{Li} &\rightarrow & P^\dag_{ij} f_{Lj} , \label{eq:f_L} \\
f_{Ri} &\rightarrow & P_{ij} f_{Rj}      , \label{eq:f_R}
\end{eqnarray}
where

\begin{equation}
P = \frac{i \omega^*}{\sqrt{3}} \left(\begin{array}{ccc}  \omega & 1      & 1 \\
                                                          1      & \omega & 1 \\
                                                          1      & 1      & \omega \\
                                \end{array} \right),
\quad \quad \omega = e^{i \frac{2 \pi}{3}}.
\label{eq:P}
\end{equation}
It is straightforward to check that this is indeed a $Z_3$ symmetry, and the proof that it is responsible 
for
the democratic form of Yukawa matrices and the right-handed neutrino mass matrix can be found in
Ref.~\cite{branco_Z3}.
Here we provide an alternative proof and a very simple understanding of this peculiar symmetry.

Let us first note that a democratic matrix $M_D$ (which represents both Yukawa matrices and the right-handed
Majorana mass matrix):
\begin{equation}
M_D = \frac{1}{3} \left(\begin{array}{ccc}  1 & 1 & 1 \\
                                            1 & 1 & 1 \\
                                            1 & 1 & 1 \\
                  \end{array} \right),
\end{equation}
can be brought to a diagonal (hierarchical) form:
\begin{equation}
M_H = \left(\begin{array}{ccc}  0 & 0 & 0 \\
                                0 & 0 & 0 \\
                                0 & 0 & 1 \\
                                \end{array} \right),
\end{equation}
by a unitary transformation
\begin{equation}
M_H = U_{\cal I} M_D U_{\cal I}^T,
\label{eq:rotation}
\end{equation}
where $U_{\cal I}$ is given in Eq.~(\ref{eq:UI3}).

It is obvious that a $Z_3$ symmetry under which the first two generations of left-handed fermions have
charge -1, the first two generations of right-handed fermions have
charge +1 (or, equivalently, charge conjugates of right-handed fermions have charge -1), and the third
generation of fermions has charge
zero uniquely specifies the $M_H$ form of Yukawa matrices and the right-handed neutrino Majorana mass matrix. 
It follows
from
the fact that any coupling involving first and/or second generation fermions is simply forbidden. Therefore the
only non-zero element of a mass matrix is the 3-3 element.
This can be written in the form of transformations~(\ref{eq:f_L}) and~(\ref{eq:f_R}) with $P$ replaced by
$P_H$:
\begin{equation}
P_H = \left(\begin{array}{ccc}  \omega & 0      & 0 \\
                                0      & \omega & 0 \\
                                0      & 0      & 1 \\
                                \end{array} \right),
\label{eq:P_H}
\end{equation}
where the subscript H indicates that it is a symmetry transformation specifying the hierarchical form of mass
matrices. Therefore we proved:
\begin{equation}
P_H M P_H = M  \quad \Leftrightarrow \quad M = \lambda M_H.
\end{equation}
Now we can rotate this result to the democratic basis:
\begin{equation}
U_{\cal I}^T P_H U_{\cal I} U_{\cal I}^T M U_{\cal I} U_{\cal I}^T P_H U_{\cal I} =
U_{\cal I}^T M U_{\cal I}  \quad \Leftrightarrow
\quad U_{\cal I}^T M U_{\cal I} = \lambda U_{\cal I}^T M_H U_{\cal I},
\end{equation}
and we obtain:
\begin{equation}
P M^\prime P = M^\prime  \quad \Leftrightarrow \quad M^\prime = \lambda M_D ,
\end{equation}
where $M^\prime  =  U_{\cal I}^T M U_{\cal I}$ and
\begin{equation}
P  = U_{\cal I}^T P_H U_{\cal I}.
\end{equation}
Inserting $P_H$ from Eq.~(\ref{eq:P_H}) and $U_{\cal I}$ from Eq.~(\ref{eq:UI3}) it is straightforward to 
find that the form of the
transformation matrix $P$ which guarantees the democratic form of all mass matrices is exactly that
of Eq.~(\ref{eq:P}).

We see that the special form of $P$ in Eq.~(\ref{eq:P}) is just a simple $Z_3$
symmetry which allows only 3-3 elements in mass matrices rotated into the democratic basis. This shows that
the approach in which the right-handed neutrino Majorana mass matrix and Yukawa matrices have in the 
leading order democratic form
is equivalent to the hierarchical approach in which the 3-3 elements in Yukawa 
matrices 
and the right-handed neutrino mass matrix dominate. The results obtained in previous sections can be 
translated into results in corresponding hierarchical models~\footnote{For discussion of a similar scenario in the 
hierarchical basis, see Ref.~\cite{Dermisek:2004tx}, 
and for more general discussion of these approaches see Ref.~\cite{Dermisek:2004mf}.}. 
Namely, the prediction for the mass of the 
lightest neutrino does not change, since mass eigenstates are not basis dependent. 
The necessary (and basis independent) requirement for achieving large mixing in this scheme is  
$M_1, M_2 < 10^{-4} M_3$, i.e., the third generation right-handed neutrino has to dominate even 
more than the third generation of quarks and charged leptons.  
Thus {\it third generation dominance} is a suitable name for this scenario.

Besides this $Z_3$ symmetry, the democratic mass matrices have also
larger symmetries, like $S_3$, for example, which can be used to specify perturbation matrices in the
process of family symmetry breaking and thus distinguish between hierarchical and
democratic starting point~\footnote{Similarly, in the hierarchical approach there are many possible symmetries 
besides our $Z_3$ which allow only the 3-3 element
of mass matrices in the leading order.}. 
However, the complicated form of the $Z_3$ symmetry in the democratic basis
compared to the simple form in the hierarchical basis further amplifies the difficulties in constructing 
specific models as discussed at the end of Sec.~\ref{sec:consequences}.

The fact that these mass matrices are motivated by a simple $Z_3$ symmetry is certainly a very pleasant
feature. Unlike $S_{3L} \times S_{3R}$, this $Z_3$ symmetry acts in the same way on all
particles in each generation. Therefore this approach can be readily embedded into GUT models, like SO(10).


\section{\label{sec:conclusions} Conclusions}

We showed that both small mixing in the quark sector and large mixing in the
lepton sector can be obtained from a simple assumption of universality of Yukawa couplings and the right-handed
neutrino Majorana mass matrix in the leading order.
We discussed conditions under which bi-large mixing
in the lepton sector is achieved with a minimal amount of fine-tuning requirements for possible models.
From knowledge of the solar and atmospheric mixing angles we determined the allowed values of $\sin \theta_{13}$.
The central values of $\sin^2 \theta_{23}$ and
$\sin^2 \theta_{12}$ predict $\sin^2 \theta_{13}$ near its present experimental upper bound,
while it can be as small as 0.008 if both $\sin^2 \theta_{23}$  and $\sin^2 \theta_{12}$ are near their $3 \sigma$ 
upper bounds. 

We showed that
this approach 
is equivalent to the hierarchical approach in which the 3-3 elements in Yukawa
matrices
and the right-handed neutrino mass matrix dominate. 
The necessary (and basis independent) requirement for achieving large mixing in this scheme is
$M_1, M_2 < 10^{-4} M_3$.
This is phenomenologically very interesting because it was found that under these conditions
the effective Majorana mass in neutrinoless double beta decay 
might be related to
the CP violating phase controlling leptogenesis~\cite{Pascoli:2003uh}. Furthermore, since the 
heaviest
right-handed
neutrino effectively decouple, this scenario might provide a natural framework for models with two right-handed 
neutrinos only~\cite{Frampton:2002qc,Raidal:2002xf,Raby:2003ay}.

The virtue of this approach is that all mass matrices are treated in the same way, providing a 
simple
framework which can be easily embedded into GUTs. 
If embedded into simple GUTs, the third generation Yukawa
coupling unification (at least approximate) is inevitable which
is theoretically very appealing and
it has interesting consequences for phenomenology~\cite{yukawa}.
Note that in this framework it is achieved without making one generation different
from others at a fundamental level.
On the other hand, the 
spectrum of the
first two generations of quarks and charged leptons crucially depends on small perturbations. 
In the neutrino sector, the heavier two neutrinos are model dependent, while the mass of the
lightest neutrino in this approach does not depend on perturbations in the leading order.
The right-handed neutrino mass scale can
be identified with the GUT scale, in which case
the mass of the lightest neutrino
is given as $(m_{top}^2/M_{GUT}) \sin^2 \theta_{23}
\sin^2 \theta_{12}$ in the limit $\sin \theta_{13} \simeq 0$.

We do not provide any understanding of the origin of universal mass matrices and their perturbations. 
It is not straightforward to construct such models with family symmetries. Nevertheless, it is worthwhile to 
look for 
alternatives: extra dimensions or composite models.
No matter what the origin is, having all three generations indistinguishable  
in leading order is certainly something one would
like to see in the fundamental theory. After all, the three generations have exactly the same quantum 
numbers in the standard model and even in simple GUT models. Why should their Yukawa couplings be so 
different?

\vspace{0.5cm}

\begin{acknowledgments}
I would like to thank S. Raby, G. Senjanovi\'c and members of the high energy theory and cosmology 
groups at UC Davis for useful comments and discussions. 
This work was supported, in part, by the U.S.\ Department of Energy, Contract
DE-FG03-91ER-40674 and the Davis Institute for High Energy Physics.
\end{acknowledgments}


\begin{thebibliography}{99}

\bibitem{Fritzsch_Xing_review}
for a review, see
H. Fritzsch and Z. Xing, Prog. Part. Nucl. Phys. {\bf 45}, 1 (2000).

\bibitem{neutrino_reviews}
for recent reviews, see
S. F. King, arXiv:hep-ph/0310204; 
A. Y. Smirnov, arXiv:hep-ph/0311259.

\bibitem{anarchy}
L. Hall, H. Murayama and N. Weiner, \prl{84}, 2572 (2000).

\bibitem{democracy}
H. Harari, H. Haut and J. Weyers, \plb{78}, 459 (1978); 
Y. Koide, \prd{39}, 1391 (1989);

\bibitem{dem_neutrino}
H. Fritzsch and Z. Xing, \plb{372}, 265 (1996); 
M. Fukugita, M. Tanimoto and T. Yanagida \prd{57}, 4429 (1998).

\bibitem{Mohapatra_Nussinov}
R. N. Mohapatra and S. Nussinov, \plb{441}, 299 (1998).


\bibitem{Dermisek:2003rw}
R.~Dermisek,
arXiv:hep-ph/0312206v1.


\bibitem{branco_efd}
E. Kh. Akhmedov, G.C. Branco, F.R. Joaquim and J.I. Silva-Marcos, \plb{498}, 237 (2001);

\bibitem{branco_Z3}
G.C. Branco and J.I. Silva-Marcos, \plb{526}, 104 (2002).

\bibitem{teshima}
T. Teshima and T. Asai, \ptp{105}, 763 (2001); 
%
T. Teshima, T. Asai and Y. Abe, \prd{66}, 093011 (2002).


\bibitem{see-saw}
M. Gell-Mann, P. Ramond and R. Slansky, in Supergravity, ed.
P. van Nieuwenhuizen and D.Z. Freedman, North-Holland, Amsterdam, 1979,
p. 315;
T. Yanagida, in Proceedings of the Workshop on the unified theory and the
baryon number of the universe, ed. O. Sawada and A. Sugamoto, KEK report No.
79-18, Tsukuba, Japan, 1979; R.~N.~Mohapatra and G.~Senjanovic, \prl{44}, 912 (1980). 


\bibitem{Gonzalez-Garcia:2003}
M.~C.~Gonzalez-Garcia and C.~Pena-Garay,
Phys.\ Rev.\ D {\bf 68}, 093003 (2003);

\bibitem{Maltoni:2003}
M.~Maltoni, T.~Schwetz, M.~A.~Tortola and J.~W.~F.~Valle,
Phys.\ Rev.\ D {\bf 68}, 113010 (2003).

\bibitem{Dermisek:2004tx}
R.~Dermisek,
arXiv:hep-ph/0406017.

\bibitem{Dermisek:2004mf}
R.~Dermisek,
arXiv:hep-ph/0409195.

\bibitem{Pascoli:2003uh}
S.~Pascoli, S.~T.~Petcov and W.~Rodejohann,
Phys.\ Rev.\ D {\bf 68}, 093007 (2003)
[arXiv:hep-ph/0302054]. 
However, see also W. Rodejohann, Eur.\ Phys.\ J.\ C {\bf 32}, 235 (2004).


\bibitem{Frampton:2002qc}
P.~H.~Frampton, S.~L.~Glashow and T.~Yanagida,
Phys.\ Lett.\ B {\bf 548}, 119 (2002)
[arXiv:hep-ph/0208157].


\bibitem{Raidal:2002xf}
M.~Raidal and A.~Strumia,
Phys.\ Lett.\ B {\bf 553}, 72 (2003)
[arXiv:hep-ph/0210021].

\bibitem{Raby:2003ay}
S.~Raby,
Phys.\ Lett.\ B {\bf 561}, 119 (2003)
[arXiv:hep-ph/0302027].


\bibitem{yukawa}
See for example recent studies,
T.~Blazek, R.~Dermisek and S.~Raby, \prl{88}, 111804 (2002);
%
ibid \prd{65}, 115004 (2002);
%
K.~Tobe and J.~D.~Wells,
Nucl.\ Phys.\ B {\bf 663}, 123 (2003);
%
D.~Auto, H.~Baer, C.~Balazs, A.~Belyaev, J.~Ferrandis and X.~Tata,
JHEP {\bf 0306}, 023 (2003);
%
R.~Dermisek, S.~Raby, L.~Roszkowski and R.~Ruiz De Austri,
JHEP {\bf 0304}, 037 (2003);
%
S.~Komine and M.~Yamaguchi, Phys.\ Rev.\ D {\bf 65}, 075013 (2002);
U. Chattopadhyay, A. Corsetti and P. Nath, \prd{66}, 035003 (2002);
S.~Profumo, Phys.\ Rev.\ D {\bf 68}, 015006 (2003);
C.~Balazs and R.~Dermisek, JHEP {\bf 0306}, 024 (2003);
C.~Pallis, Nucl.\ Phys.\ B {\bf 678}, 398 (2004);
and references therein.


\end{thebibliography}

\end{document}